\documentclass[aps,prl,twocolumn,preprintnumbers]{revtex4-2}
\usepackage{amsmath, mathrsfs, amssymb,amsfonts,amsthm,graphicx, epsf, dcolumn, yfonts, subfigure}
\usepackage[hyperfootnotes=true]{hyperref}
\usepackage{color}
\usepackage{slashed}
\usepackage{setspace}
\usepackage{cancel}
\usepackage{wasysym}
\usepackage{float}
\pdfoutput=1
 \newcommand{\be}{\begin{equation}}
 \newcommand{\ee}{\end{equation}}
 \newcommand{\bea}{\begin{eqnarray}}
 \newcommand{\eea}{\end{eqnarray}}

\newcommand{\beq}{\begin{equation}}
\newcommand{\eeq}{\end{equation}}

\renewcommand*{\thefootnote}{\fnsymbol{footnote}}

\begin{document}

\preprint{IFT-UAM/CSIC-24-94}

\title{Quantum inequalities for quantum black holes}
\author{Antonia M. Frassino,$^{1,2}$ Robie A. Hennigar,$^{2}$ Juan F. Pedraza$^{3}$ and Andrew Svesko$^{4}$}
\affiliation{\vspace{1mm}
$^1$Departamento de F\'{i}sica y Matem\'{a}ticas, Universidad de Alcal\'{a}, Campus Universitario, Alcal\'a de Henares, 28805 Madrid, Spain\\
$^2$Departament de Física Quàntica i Astrofísica and
  Institut de Ciències del Cosmos, Universitat de Barcelona, 08028 Barcelona, Spain\\
$^3$Instituto de F\'isica Te\'orica UAM/CSIC, Cantoblanco, 28049 Madrid, Spain\\
$^4$Department of Mathematics, King’s College London, Strand, London, WC2R 2LS, UK}

\begin{abstract}\vspace{-2mm}

 \noindent We formulate spacetime inequalities applicable to quantum-corrected black holes to all orders of backreaction in semiclassical gravity. Namely, we propose refined versions of the quantum Penrose and reverse isoperimetric inequalities, valid for all known three-dimensional asymptotically anti-de Sitter quantum black holes. Previous proposals of the quantum Penrose inequality apply in higher dimensions but fail when applied in three dimensions beyond the perturbative regime. Our quantum Penrose inequality, valid in three dimensions, holds at all orders of backreaction. This suggests cosmic censorship must exist in non-perturbative semiclassical gravity. Our quantum reverse isoperimetric inequality implies a maximum entropy state for quantum black holes at fixed volume. 

% \noindent We formulate generalizations of classical spacetime inequalities applicable to quantum-corrected black holes at all orders of semiclassical backreaction due to quantum matter. Specifically, we propose refined versions of the quantum Penrose and reverse isoperimetric inequalities, showcasing their validity across all known three-dimensional asymptotically anti-de Sitter quantum black holes. Previous proposals of the quantum Penrose inequality only hold for perturbative quantum effects in higher dimensions. We propose a quantum Penrose inequality in three-dimensions valid at all orders of backreaction. This suggests quantum cosmic censorship should exist in semiclassical gravity. Our conjectured quantum reverse isoperimetric inequality, which employs Casimir-subtracted thermodynamic volume, implies a maximum entropy state for quantum black holes.
 
 %, clarifying earlier findings in the literature. 

\end{abstract}

\renewcommand*{\thefootnote}{\arabic{footnote}}
\setcounter{footnote}{0}

\maketitle

\noindent \textbf{Introduction.} Black holes play a key role in understanding the relation between geometry and matter. A potent example is given by the conjectured Penrose inequality (PI) \cite{Penrose:1973um}, which, roughly, quantifies the mass of a spacetime in terms of the black holes it contains. More precisely, assuming singularities are hidden behind event horizons and collapsing matter settles to a Kerr black hole, then the total %Arnowitt-Deser-Misner (ADM) 
mass $M_{\text{ADM}}$ for a four-dimensional asymptotically flat spacetime with a marginally trapped surface $\sigma$ is bounded below by the area $A[\sigma]$, 
\beq G_{4}M_{\text{ADM}}\geq\sqrt{\frac{A[\sigma]}{16\pi}}\;.
\label{eq:PenroseinqOG}\eeq
The bound is saturated for the Schwarzschild black hole while adding rotation gives a strict inequality. The conjecture has been proven in special cases \cite{Huisken01,Bray01,Bray:2007opu}, and may be generalized to higher-dimensions \cite{Bray:2007opu} and asymptotically anti-de Sitter (AdS) spaces \cite{Itkin:2011ph,Folkestad:2022dse}. Further, given the thermal nature of black holes, where entropy is proportional to the area of their event horizon \cite{Bekenstein:1972tm,Bekenstein:1973ur,Hawking:1974sw}, the inequality (\ref{eq:PenroseinqOG}) can be reinterpreted as an entropy bound. 

Black holes in AdS are proposed to obey another bound, the reverse isoperimetric inequality (RII) \cite{Cvetic:2010jb}
\be 
\mathcal{R} \equiv \left(\frac{(D-1) V_{\rm th}}{\Omega_{D-2}} \right)^{\frac{1}{D-1}} \left(\frac{\Omega_{D-2}}{A_{\text{BH}}}\right)^{\frac{1}{D-2}}\geq1\;.
\label{eq:classRII}\ee
Here $\Omega_{D-2}$ is the volume of a unit $(D-2)$ sphere, of $D$-dimensional AdS, $A_{\text{BH}}$ is the area of the black hole horizon, and $V_{\rm th}$ is the `thermodynamic volume'. This inequality is motivated by the framework of \emph{extended} black hole thermodynamics \cite{Kastor:2009wy,Dolan:2010ha}, where the cosmological constant is treated as a variable pressure. In this context, the inequality says a black hole with fixed thermodynamic volume has an entropy no larger than Schwarzschild-AdS of the same volume.  While lacking a generic proof, there are no known counterexamples to (\ref{eq:classRII}), except possibly the charged  Ba\~{n}ados-Teitelboim-Zanelli (BTZ) black hole \cite{Martinez:1999qi}.\footnote{For $D\geq4$, RII is saturated for Schwarzschild-AdS, with strict inequality for rotating black holes. `Ultraspinning' black holes \cite{Klemm:2014rda,Hennigar:2014cfa,Hennigar:2015cja}, were initially thought to violate RII, i.e.,  $\mathcal{R}<1$, though was called into question \cite{Appels:2019vow}. For $D=3$, both the static and rotating BTZ black hole saturate (\ref{eq:classRII}). Electrically charged BTZ, however, violates the RII \cite{Frassino:2015oca}. `Exotic' BTZ black holes \cite{Carlip:1991zk,Carlip:1994hq,Townsend:2013ela} %were also argued to violate 
have been suggested to potentially violate
(\ref{eq:classRII}) \cite{Frassino:2019fgr,Cong:2019bud}, however, when the event horizon area is replaced by the correct form of the horizon entropy, the exotic BTZ black holes are found to obey $\mathcal{R}\geq1$ \cite{Johnson:2019wcq}.} 
Refined generalizations of (\ref{eq:classRII}), inspired by the Penrose inequality (\ref{eq:PenroseinqOG}), have been conjectured and broadly tested \cite{Amo:2023bbo}.

The conjectured inequalities (\ref{eq:PenroseinqOG}) and (\ref{eq:classRII}) are classical.
%Matter, however, is fundamentally quantum. 
It is natural to wonder how they fair under quantum effects. The PI is known to be violated \cite{Bousso:2019var,Bousso:2019bkg} for quantum matter coupled to classical gravity. As such, a quantum Penrose inequality (QPI) was conjectured, where, in the spirit of semiclassical generalizations of established classical principles \cite{Wall:2010jtc,Bousso:2015mna,Bousso:2015eda,Bousso:2015wca,Balakrishnan:2017bjg}, area $A[\sigma]$ is replaced by the generalized entropy \cite{Bekenstein:1974ax} associated to a quantum trapped surface. Evidence gathered thus far suggests the QPI is obeyed for small perturbative backreaction. Meanwhile, the status of RII (\ref{eq:classRII}) under backreaction effects is unclear, though preliminary evidence in favor of a semiclassical generalization was offered in \cite{Frassino:2022zaz}. A complete assessment of either quantum inequality requires solving the semiclassical Einstein equations, an open problem in $D\geq3$ spacetime dimensions.

 Here we propose and test quantum Penrose and reverse isoperimetric inequalities using exact black holes in semiclassical gravity. 
 %including all orders of backreaction due to a large number of quantum fields.
 Our tests rely on braneworld holography~\cite{deHaro:2000wj}. In this framework a $(D-1)$-dimensional end-of-the-world brane is coupled to general relativity in a $D$-dimensional asymptotically AdS space, which has a dual description as a conformal field theory (CFT) living on the AdS boundary. As in holographic regularization \cite{deHaro:2000vlm}, the brane renders the (on-shell) bulk action finite. 
%by integrating out bulk degrees of freedom to the brane.
A higher curvature gravity theory is induced on the brane coupled to a CFT with large central charge and an ultraviolet cutoff. 
%that backreacts on the brane geometry. 
Importantly, with this formalism quantum-corrected black holes in \emph{three dimensions} can be exactly constructed to all orders of backreaction \cite{Emparan:2002px,Reviewtoappear}. 

%Thus, classical black hole solutions to the bulk theory map to quantum-corrected black holes on the brane, to  all orders of backreaction \cite{Emparan:2002px}. 

\noindent \textbf{Quantum inequalities.} \emph{Quantum Penrose inequality.} 
%Classically, the Penrose inequality for asymptotically AdS spacetimes of mass $M$ and dimension $D\geq3$ obeying the Einstein equations  and the dominant energy condition is \cite{Itkin:2011ph,Folkestad:2022dse},
Classically, the Penrose inequality for $D\geq4$ AdS spacetimes is as follows. Assuming cosmic censorship and collapsing matter settles to Kerr-AdS, then \cite{Itkin:2011ph,Folkestad:2022dse}
\beq \frac{16\pi G_{D}M_{\text{AMD}}}{(D-2)\Omega_{D-2}}\geq \left(\frac{A[\sigma]}{\Omega_{D-2}}\right)^{\hspace{-1mm}\frac{D-3}{D-2}}+\ell^{-2}_{D}\left(\frac{A[\sigma]}{\Omega_{D-2}}\right)^{\hspace{-1mm}\frac{D-1}{D-2}}.
\label{eq:classAdSPI}\eeq
Here $G_{D}$ and $\ell_{D}$ denote the $D$-dimensional Newton's constant and curvature scale, respectively, $\sigma$ is a (outermost) marginally trapped surface, 
and 
%$\Omega_{k}$ is the volume of a $(D-2)$-dimensional unit sphere ($k=1$), plane ($k=0$), or hyperbolic space ($k=-1$); we will be interested in the spherical case where, 
$\Omega_{n}\equiv 2\pi^{(n+1)/2}/\Gamma[(n+1)/2]$ is the volume of a unit $n$-sphere. The inequality (\ref{eq:classAdSPI}) assumes spherical symmetry, however, is easily generalized to planar or hyperbolic symmetry. As written \cite{Folkestad:2022dse}, one assumes the Ashtekar-Magnon-Das (AMD)~\cite{Ashtekar:1984zz,Ashtekar:1999jx,Gibbons:2004ai,Hollands:2005wt} convention for mass in AdS (for $D>3$), where global AdS has vanishing mass. In \cite{Itkin:2011ph}, the local counterterm prescription of AdS mass \cite{Balasubramanian:1999re,Emparan:1999pm} is used, such that the left side of (\ref{eq:classAdSPI}) is shifted by the non-vanishing mass of empty AdS, i.e., the Casimir energy $M_{\text{cas}}$.
%(equaling the Casimir energy of the dual CFT, when there exists a holographic description).
We will assign zero mass for global AdS for $D\geq3$.
%, such that the blackening factor for the AdS-Schwarzschild black hole is
%\beq f(r)=1-\frac{16\pi G_{D}M_{\text{AMD}}}{(D-2)\Omega_{D-2}r^{D-3}}+\frac{r^{2}}{\ell_{D}^{2}}.\label{eq:blackfact}\eeq
Inequality (\ref{eq:classAdSPI}) saturates for AdS-Schwarzschild and is strict otherwise. 
%For $\ell_{D}\to\infty$ and $D=4$, Eq. (\ref{eq:classAdSPI}) reduces to (\ref{eq:PenroseinqOG}). 
%to the flat space inequality 

% There is a distinction between cases $D>3$ and $D=3$. For $D>3$, the inequality (\ref{eq:classAdSPI}) is saturated for the AdS-Schwarzschild black hole, and strict inequality for Kerr-AdS. In $D=3$, however, the static BTZ black hole obeys
% \beq 8G_{3}M_{\text{BTZ}}=L_{3}^{-2}\left(\frac{2\pi r_{+}}{2\pi}\right)^{2}\;,\eeq
% for horizon radius $r_{+}$, which is not equal to (\ref{eq:classAdSPI}) with $D=3$. 

The inequality (\ref{eq:classAdSPI}) can be violated by semiclassical quantum effects. Following \cite{Bousso:2019var}, this can be demonstrated by considering, e.g., quantum fields in the Boulware vacuum. Perturbatively, negative energy density due to the fields near the horizon leads to a negative contribution to the mass such that (\ref{eq:classAdSPI}) is violated. Given that any violation of the Penrose inequality implies a failure of (weak) cosmic censorship, this motivates the question of whether a semiclassical generalization of  (\ref{eq:classAdSPI}) exists. 

A proposed quantum Penrose inequality for $D\geq4$ asymptotically AdS spacetimes is \cite{Bousso:2019bkg}
\beq
\begin{split} 
\hspace{-4mm}\frac{16\pi \mathcal{G}_{D}M_{\text{AMD}}}{(D-2)\Omega_{D-2}}&\geq \left(\frac{4\mathcal{G}_{D}S_{\text{gen}}}{\Omega_{D-2}}\right)^{\hspace{-1mm}\frac{D-3}{D-2}}\hspace{-2mm}+\ell^{-2}_{D}\left(\frac{4\mathcal{G}_{D}S_{\text{gen}}}{\Omega_{D-2}}\right)^{\hspace{-1mm}\frac{D-1}{D-2}}
\end{split}
\label{eq:qAdSPI}\eeq
where area has been replaced by generalized entropy
\beq S_{\text{gen}}=\frac{A[\Sigma]}{4\mathcal{G}_{D}}+S_{\text{vN}}^{\text{mat}}+S_{\text{Wald}}\;.\label{eq:sgengen}\eeq
Here $A[\Sigma]$ is the codimension-2 area of a Cauchy-splitting surface $\Sigma$, $S_{\text{vN}}^{\text{mat}}\equiv-\text{tr}(\rho\log\rho)$ is the von Neumann entropy of state $\rho$ of quantum fields living on the classical background confined to one side of $\Sigma$.
%Generally, the matter entropy is divergent in the ultraviolet (UV) due to vacuum entanglement across $\Sigma$, with a leading order contribution of the form $A[\Sigma]/\epsilon^{D-2}$, for UV regulator $\epsilon$, as well as subleading terms in $\epsilon$. 
The gravitational area in (\ref{eq:qAdSPI}), with renormalized Newton's constant $\mathcal{G}_{D}$, regularizes the leading area divergence of the matter entropy while the subleading divergences are regulated via the  Wald entropy \cite{Wald:1993nt} accounting for higher-derivative gravitational couplings. Thus, $S_{\text{gen}}$ is finite in the ultraviolet (UV) \cite{Susskind:1994sm,Solodukhin:2011gn,Cooperman:2013iqr}. 
Technically, moreover, the generalized entropy in (\ref{eq:sgengen}) is evaluated over a \emph{quantum} marginally trapped surface \cite{Bousso:2019var,Bousso:2019bkg}, i.e., a surface for which the (outer) inner \emph{quantum} expansion of future-directed null-rays orthogonal to it is (vanishing) non-positive \cite{Bousso:2015mna,Bousso:2015eda}.
%\footnote{The quantum expansion $\Theta$ of a surface $\mu$ is the semiclassical generalization of classical expansion, formally defined as the functional derivative of the generalized entropy with respect to a function specifying the affine location of $\mu$ and nearby surfaces along a congruence of null geodesics orthogonal to $\mu$ \cite{Bousso:2015mna}.} 

%A version of the quantum inequality (\ref{eq:qAdSPI}) was previously proposed in \cite{Bousso:2019bkg}. Our proposal differs in one essential way. We explicitly subtract the Casimir energy  $M_{\text{cas}}$ due to quantum matter fields in vacuum. Our motivation for this subtraction is because without it the quantum Penrose inequality will be violated when backreaction effects are large. 
%If the goal is to develop a notion of cosmic censorship for semiclassical gravity, such violations should be excluded.
%We find evidence for this below. 

%Given the connection between the Penrose inequality and cosmic censorship, 
While the QPI (\ref{eq:qAdSPI}) has been demonstrated to hold perturbatively, it is worth testing its validity when backreaction effects are large. This requires a self-consistent solution to the semiclassical Einstein equations in $D\geq4$, which is lacking. This motivates us to descend to $D=3$, where the backreaction problem can be solved exactly using braneworld holography. We find that naive application of (\ref{eq:qAdSPI}) in $D=3$ results in violations beyond the perturbative regime. Instead, we propose,
%the following quantum Penrose inequality for AdS$_{3}$
%and we propose 
\beq 8\pi\mathcal{G}_{3}M_{\text{AMD}}\geq\ell_{3}^{-2}\left(\frac{4\mathcal{G}_{3}S_{\text{gen}}}{2\pi}\right)^{2}\;.\label{eq:AdS3qPeninq}\eeq
Below we find evidence this inequality holds at all orders of backreaction, a consequence of $M_{\text{AMD}}$ subtracting the Casimir energy of backreacting quantum fields.

Inequality (\ref{eq:AdS3qPeninq}) is visibly different from its $D\geq4$ counterpart (\ref{eq:qAdSPI}). This is unsurprising since the classical Penrose inequality in $D=3$ is more subtle than in higher dimensions.
%Firstly, this is because in $D=3$ the AMD prescription is not applicable since the Weyl tensor vanishes. In this article we assign zero mass for global AdS for $D\geq3$.
Setting $D=3$ in (\ref{eq:classAdSPI}), the first term on the right-hand side reduces to unity. While the resulting inequality appears saturated for the static BTZ black hole (a useful guiding principle in $D\geq4$), there is no known derivation of the inequality in this form.\footnote{The Penrose-like inequality derived in \cite{Bengtsson:2016dac}, following \cite{Mars:2012ym}, results from integrating the expansion of a collapsing null shell.} Conceptually, moreover, unlike in $D\geq4$, black holes in AdS$_{3}$ formed under collapse cannot have arbitrarily small mass -- a consequence of a gap in the mass spectrum between empty AdS$_{3}$ and the BTZ black hole \cite{Banados:1992gq}. Regarding the Penrose inequality, this means the mass of an asymptotically AdS$_{3}$ initial data with a marginally trapped surface is not expected to go below the mass gap. 
%nor should we expect the static BTZ black hole to saturate the Penrose inequality.
Consequently, we propose that the classical PI for AdS$_{3}$ should be substantively different from (\ref{eq:classAdSPI}) in that the first term on the right-hand side is not present. This follows from the classical limit of our proposed quantum inequality (\ref{eq:AdS3qPeninq}). %implies a classical Penrose inequality in AdS$_{3}$.

\noindent \emph{Quantum reverse isoperimetric inequality.} Euler's theorem of homogeneous functions implies black holes with a non-zero cosmological constant $\Lambda_{D}$ obey \cite{Kastor:2009wy}
\beq (D-3)G_{D}M=(D-2)TS-2P_{D}V+...\;,\label{eq:extSmarr}\eeq
for temperature $T$, Bekenstein-Hawking entropy $S=A_{\text{BH}}/4G_{D}$, and the ellipsis refers to other possible conserved charges multiplied by associated potentials. Further, $P_{D}\equiv-\Lambda_{D}/8\pi G_{D}$ is a pressure and $V\equiv(\frac{\partial M}{\partial P_{D}})_{S,...}$ is its conjugate (`thermodynamic') volume. For vanishing cosmological constant  (\ref{eq:extSmarr}) reduces to the Smarr formula, but for $\Lambda_{D}\neq0$ the $P-V$ term is required for consistency. Treating $P_{D}$ as a dynamical variable leads to an extended framework of black hole thermodynamics.%\footnote{In classical gravity, the thermodynamic volume has a geometric definition in terms of Komar integrals. In this case it is not necessary to vary the cosmological constant to define and study the thermodynamic volume.} 

%The interpretation of the thermodynamic volume remains largely mysterious. For simple cases it coincides with the geometric volume occupied by the black hole ---the amount of spacetime volume excluded by the black hole horizon--- but in general it differs \cite{Dolan:2010ha,Cvetic:2010jb,Johnson:2014xza}.
%\footnote{Alternatively, it has been suggested the term `volume' should be dispensed in favor of a more general notion of gravitational tension \cite{Armas:2015qsv}.} 
%Nonetheless, it is the thermodynamic volume that obeys the reverse isoperimetric inequality (\ref{eq:classRII}). Given that all known violations of RII are for black holes that are thermally unstable in fixed volume sector \cite{Johnson:2019mdp}, it is clear $V_{\text{th}}$ plays a key role in understanding black hole thermodynamics. Moreover, the thermodynamic volume has a zero-point ambiguity tied to the zero-point ambiguity with defining mass (see supplemental material of~\cite{Amo:2023bbo}). The RII holds when mass is defined according to the AMD convention, i.e., subtracting off Casimir energy.  

The interpretation of the thermodynamic volume remains largely mysterious. For simple cases it coincides with the geometric volume occupied by the black hole ---the amount of spacetime volume excluded by the black hole horizon--- but in general it differs~\cite{Dolan:2010ha,Cvetic:2010jb,Johnson:2014xza}. The thermodynamic volume has a zero-point ambiguity tied to the zero-point ambiguity with defining mass (see supplemental material of~\cite{Amo:2023bbo}). The RII holds when mass is defined according to the AMD convention, i.e., subtracting off Casimir energy. There are few examples of black holes that violate the RII and there are ambiguities associated with the thermodynamics of all such cases. Universally, however, all known violations of the RII are thermally unstable black holes~\cite{Johnson:2019mdp}, proving $V$ plays a key role in understanding black hole thermodynamics.

%\AS{Say something about nature abhors `superentropic' black holes?}

With this in mind, we propose the natural quantum generalization of the classical RII (\ref{eq:classRII}) be
\be 
\mathcal{R}_{\text{Q}} \equiv \left(\frac{(D-1) V_{\rm th}}{\Omega_{D-2}} \right)^{\frac{1}{D-1}} \left(\frac{\Omega_{D-2}}{4\mathcal{G}_{D}S_{\text{gen}}}\right)^{\frac{1}{D-2}}\geq1\;.
\label{eq:qRII}\ee
Akin to the quantum Penrose inequality, classical area has been replaced for generalized entropy (\ref{eq:sgengen}), and $V_{\text{th}}$ is the Casimir-subtracted thermodynamic volume,
\beq V_{\text{th}}\equiv V-V_{\text{cas}}\;,\quad V_{\text{cas}}\equiv\left(\frac{\partial M_{\text{cas}}}{\partial P_{D}}\right)_{\hspace{-1mm} S_{\text{gen}},...}\;.\label{eq:cassubvol}\eeq
In analogy with Casimir energy $M_{\text{cas}}$, $V_{\text{cas}}$ is the thermodynamic volume assigned to empty AdS space --- it will be nonzero precisely when the Casimir energy is nonzero. 
%Below we find evidence $\mathcal{R}_{\text{Q}}\geq1$ is robust against backreaction at all orders for thermally stable black holes. 

%\AS{ Can we show violations of classical RII event for perturbative quantum effects similar to classical Penrose inequality? }

\noindent \textbf{Evidence from quantum black holes.} We test our quantum inequalities for static and rotating quantum BTZ (qBTZ) black holes \cite{Emparan:2020znc,Climent:2024nuj,Feng:2024uia}. Each example arises when an $\text{AdS}_{3}$ end-of-the-world brane \cite{Karch:2000ct} intersects an appropriate AdS$_{4}$ C-metric black hole horizon \cite{Emparan:1999wa,Emparan:1999fd}. By braneworld holography, the geometry and thermodynamics of qBTZ are known analytically and the inequalities may be tested at all orders of backreaction. 

\noindent \emph{Exact description of quantum black holes.} Let us first summarize key features of quantum AdS$_{3}$ black holes. For example, the metric of the static, neutral quantum BTZ black hole \cite{Emparan:2020znc} of mass $M$ is (see the supplemental material for descriptions of charged and rotating metrics) 
% material for charged and rotating descriptions)
%follows from introducing an $\text{AdS}_{3}$   end-of-the-world brane \cite{Karch:2000ct} inside an asymptotically  AdS$_{4}$ space described by the C-metric \cite{Emparan:1999wa,Emparan:1999fd}. The brane intersects the $\text{AdS}_{4}$ black hole horizon such that it localizes on the brane.  By braneworld holography, the backreacted geometry and horizon thermodynamics of the qBTZ are known analytically. 
\beq
\begin{split}
&ds^2=-f(r)dt^2+\frac{dr^2}{f(r)}+r^2d\phi^2\,,\\
&f(r)=\frac{r^2}{\ell_3^2}-8\mathcal{G}_3M-\frac{\ell F(M)}{r}\,.
\end{split}
\label{eq:qBTZmain}\eeq
Here, 
$\ell$ is the UV cutoff length scale, 
  $\mathcal{G}_3=G_{3}/\sqrt{1+(\ell/\ell_{3})^{2}}$ is the `renormalized' Newton's constant, and $F(M)$ is a positive function of the mass, the details of which are unnecessary for our purposes. The $\text{AdS}_{3}$ length $\ell_{3}$ is related to the brane cosmological constant (see supplemental material).
% \beq \label{lambda} \Lambda_3 \equiv - \frac{1}{L_3^2} 
% = -  2 \left ( \frac{1}{\ell^2} + \frac{1}{\ell_3^2} - \frac{1}{\ell} \sqrt{\frac{1}{\ell^2} + \frac{1}{\ell_3^2} } \right)\,.
% \eeq
The mass $M$ includes the Casimir energy of the CFT, $M=M_{\text{AMD}}+M_{\text{cas}}$. 
%\rah{The AMD mass does not exist in $D = 3$, but if the intention here is just to mean that $M = 0$ for \textit{global} AdS, then I understand what is meant.} \AS{Yes, that is the intention.}

The metric (\ref{eq:qBTZmain})  may be understood as a quantum black hole in that it is a solution to the induced semiclassical theory on the brane at all orders in quantum backreaction. From this perspective, the parameter $\ell$ controls the strength of backreaction due to the cutoff $\text{CFT}_{3}$. For small backreaction, $\ell/\ell_{3}\ll1$, then $L_{3}^{2}\approx \ell_{3}^{2}$ while $2c_{3}G_{3} \approx\ell$ for central charge $c_{3}\gg1$. Vanishing backreaction occurs when $\ell\to0$, for fixed $c_{3}$, where gravity becomes weak on the brane. Since, $\ell\approx 2c_{3}L_{\text{P}}\gg L_{\text{P}}$ for Planck length $L_{\text{P}}=G_{3}$ (with $\hbar=1$), the $\sim\ell/r$ quantum correction in the metric (\ref{eq:qBTZmain})  is not Planckian.

The thermodynamics of the quantum BTZ black hole is inherited from the AdS$_{4}$ bulk black hole thermodynamics. The mass $M$, temperature $T$ and entropy $S$ of the classical bulk black hole are \cite{Emparan:1999fd,Emparan:2020znc} 
%\rah{The mass can also be consistently computed using only the brane description --- should that be emphasized?} \AS{I think it is fine if we don't mention it.}
\bea
M&=&\frac{\sqrt{1+\nu^{2}}}{2G_3}\frac{z^2(1-\nu z^3)(1+\nu z)}{(1+3z^2+2\nu z^3)^2}\,, \label{eq:qbtzthermo}\\
T&=&\frac{1}{2\pi\ell_3}\frac{z(2+3\nu z+\nu z^3)}{1+3z^2+2\nu z^3}\,, \label{eq:temp}\\
S&=&\frac{\pi \ell_3\sqrt{1+\nu^{2}}}{G_3}\frac{z}{1+3z^2+2\nu z^3}\,,
\label{eq:genentropy}
\eea
where $z\!\equiv \!\ell_3/(r_{+}x_1)$ for horizon radius $r_{+}$, and $\nu \! \equiv \!\ell / \ell_3$ both have range $[0,\infty)$ ($x_{1}$ is a geometric parameter of the bulk C-metric). Each quantity may be derived by identifying the bulk on-shell Euclidean action with the canonical free energy~\cite{Kudoh:2004ub}. 
%Previous work \cite{Appels:2016uha,Anabalon:2018ydc} examined accelerating black hole thermodynamics but not in the presence of a brane.
On the brane, the quantum black hole has the same temperature $T$, while the four-dimensional  Bekenstein-Hawking entropy $S$ is identified as the three-dimensional generalized entropy,  $S\equiv S_{\text{gen}}$ \cite{Emparan:2006ni,Emparan:2020znc}, accounting for both (higher-curvature) gravitational and semiclassical matter entropy as in (\ref{eq:sgengen}). Thence, for $\ell$ and $\ell_3$ fixed, the first law of thermodynamics  is
\beq dM=TdS_{\text{gen}}\,,\label{eq:qbtzfirstlaw}\eeq
and is valid to all orders in backreaction. 
%and where the qBTZ mass is identified as $M$. Classical entropy being replaced by the generalized entropy in the first law also occurs for two-dimensional semiclassical black holes \cite{Pedraza:2021cvx,Svesko:2022txo}.

\noindent \emph{Quantum Penrose inequality.} Let us now test our proposal (\ref{eq:AdS3qPeninq}) for quantum AdS$_{3}$ black holes. First note the mass $M$ in the static geometry (\ref{eq:qBTZmain}) has a finite range, including negative masses corresponding to quantum dressed AdS$_{3}$ conical singularities. The lower bound on $M$ is the zero-point energy of the UV cutoff CFT
\be 
M_{\rm cas} = - \frac{\sqrt{1+\nu^2}}{8 G_3} =  - \frac{1}{8 \mathcal{G}_3 }\, .
\label{eq:CasmirMqbtz}\ee
 In the limit of vanishing backreaction $\nu\to0$, $M_{\text{cas}}$ agrees with the Casimir energy found using the local counterterm prescription of AdS mass \cite{Balasubramanian:1999re}. Incidentally, $M_{\text{cas}}$ coincides with the $z\to\infty$ limit of mass $M$ (\ref{eq:qbtzthermo}).

Subtracting the Casimir energy (\ref{eq:CasmirMqbtz}) from the mass $M$ of any of the AdS$_{3}$ quantum black holes (see supplemental material for explicit expressions), we find our proposed inequality (\ref{eq:AdS3qPeninq}), with $M_{\text{AMD}}=M-M_{\text{cas}}$ holds for all $\nu$.\footnote{The same inequality with $\mathcal{G}_{3}$ replaced by $G_{3}$ also holds.}
%for all $\nu$,
% \be 
% 8 \mathcal{G}_3 \left( M - M_{\rm cas} \right)  -  \frac{1}{\ell_3^2} \left(\frac{4\mathcal{G}_3 S_{\rm gen}}{ 2\pi} \right)^2 \ge 0 \, .
% \label{eq:fundineq}\ee
For $\nu\neq0$, (\ref{eq:AdS3qPeninq}) is a strict inequality, at least for black holes subject to positive temperature and entropy, and real values of angular momentum and velocity.  

Inequality (\ref{eq:AdS3qPeninq}) is trivially saturated for static qBTZ in the large-$z$ limit, when the black hole shrinks to arbitrarily small size and $S_{\text{gen}}\to0$. By contrast, when $\nu=0$, the resulting classical Penrose inequality is not saturated for static BTZ. This suggests saturation of the Penrose inequality is linked to the existence of a mass gap in the black hole spectrum. The reasoning is as follows. The quantum BTZ solution represents a family of black holes with a continuous spectrum -- the classically naked conical singularities resulting in the gap are shrouded by a horizon induced by backreaction. Thus, quantum effects shrink the mass gap to zero, and saturation of the quantum Penrose inequality (\ref{eq:AdS3qPeninq}) reflects that quantum effects allow for the formation of black holes with masses disallowed classically. This observation is consistent with the classical Penrose inequality for $D\geq4$ (\ref{eq:classAdSPI}), where there is no gap between AdS black holes and empty AdS.

%Adding and subtracting $8\mathcal{G}_{3}M_{\text{cas}}$ and substituting $M=M_{\text{AMD}}+M_{\text{cas}}$ yields the proposed inequality (\ref{eq:qAdSPI}) for $D=3$, recovering the classical inequality (\ref{eq:classAdSPI}) when $\nu=0$. 

%The validity of this inequality can be directly confirmed in Mathematica, using the ``Reduce'' function combined with the physicality constraints. In the case of the rotating solution, I have checked the inequality on the parameter space where $\nu < z < \nu^{-1/3}$ (subject also to positive temperature, entropy, and real values of $\Omega$ and $J$). 

% \AS{comments.} The connection between the above expression and the traditional form of the Penrose inequality in AdS needs to be clarified. (Since AdS$_3$ is special in this regard). Note that for the classical, static, and uncharged BTZ metric the following identity holds:
% \be 
% 8 G_3 M = \frac{1}{\ell_3^2} \left(\frac{2 G_3 S}{ \pi} \right)^2 \, .
% \ee

\noindent \emph{Quantum reverse isoperimetric inequality.} Thermal variables of quantum BTZ obey the Smarr relation \cite{Frassino:2022zaz}
\be 
0=TS_{\text{gen}}-2P_3V_3+...\,,
\label{eq:3dsmarr}\ee
where $P_3=-\Lambda_3/(8\pi G_{3})$ is the pressure with conjugate volume $V_{3}$
%and $\mu_{3}$ is the chemical potential conjugate to  $c_{3}$  
(see \cite{Frassino:2022zaz} and supplemental material for exact expressions).
 %Unlike higher-dimensional Smarr formulae, the mass is absent from the three-dimensional Smarr relation (\ref{eq:3dsmarr}) since $G_3 M$ has vanishing scaling dimension, as with the classical BTZ black hole \cite{Frassino:2015oca, Frassino:2019fgr}.
 Treating $P_{3}$ as a dynamical variable is natural in the context of braneworld holography \cite{Frassino:2022zaz}. Standard thermodynamics of classical bulk black holes including work done by the brane maps to extended thermodynamics of quantum black holes, e.g., dynamical pressure is dual to variable brane tension. 

 Let us test our quantum RII (\ref{eq:qRII}). Consider the ratio
% Let us begin by reviewing the most relevant features of the reverse isoperimetric inequality in classical general relativity. The inequality was originally proposed in~\cite{} and asserts that the isoperimetric ratio
% \be 
% \mathcal{R} \equiv \left(\frac{(D-1) V_{\rm th}}{\Omega_{D-2}} \right)^{\frac{1}{D-1}} \left(\frac{\Omega_{D-2}}{A}\right)^{\frac{1}{D-2}}
% \ee
% always satisfies
% \be 
% \mathcal{R}  \ge 1 \, .
% \ee
% Here $\Omega_{D-2}$ is the volume of a unit $(D-2)$ sphere, $A$ is the area of the black hole horizon, while $V_{\rm th}$ is the thermodynamic volume. Recall that the thermodynamic volume has a zero-point ambiguity that is directly tied to the zero-point ambiguity associated with defining the mass.\footnote{This is reviewed in the supplemental material of~\cite{Amo:2023bbo}.} Specifically, the claim of~\cite{Cvetic:2010jb} was that the reverse isoperimetric inequality holds when the mass is defined according to Ashtekar-Magnon-Das (AMD) prescription~\cite{Ashtekar:1984zz, Ashtekar:1999jx, Gibbons:2004ai}. In the parlance of the AdS/CFT this corresponds to subtracting off the Casimir energy contributions to the mass. 
\be 
\mathcal{R}_\text{Q} \equiv \left(\frac{2V_{\rm th}}{2\pi} \right)^{1/2} \left( \frac{2\pi }{4 \mathcal{G}_3 S_{\rm gen}} \right),\label{eq:qratio}\ee
for Casimir-subtracted volume (\ref{eq:cassubvol})
\be 
V_{\rm cas} = - \frac{\pi \nu^2 \ell_3^2 }{2} \, .
\label{eq:casvol}\ee
 Incidentally, as with Casimir mass, $V_{\text{cas}}$  coincides with the $z \to \infty$ limit of $V_{3}$ for the neutral, static qBTZ (\ref{eq:qBTZmain}). %For charged or rotating qBTZ solutions,  naively taking $z\to\infty$ at fixed values of charge or angular momentum takes us outside the allowed parameter ranges of the solutions, passing through naked singularities. We thus take the Casimir volume for all solutions we study to be (\ref{eq:casvol}).
 
\begin{figure}[t]
\centerline{\includegraphics[width=0.42\textwidth]{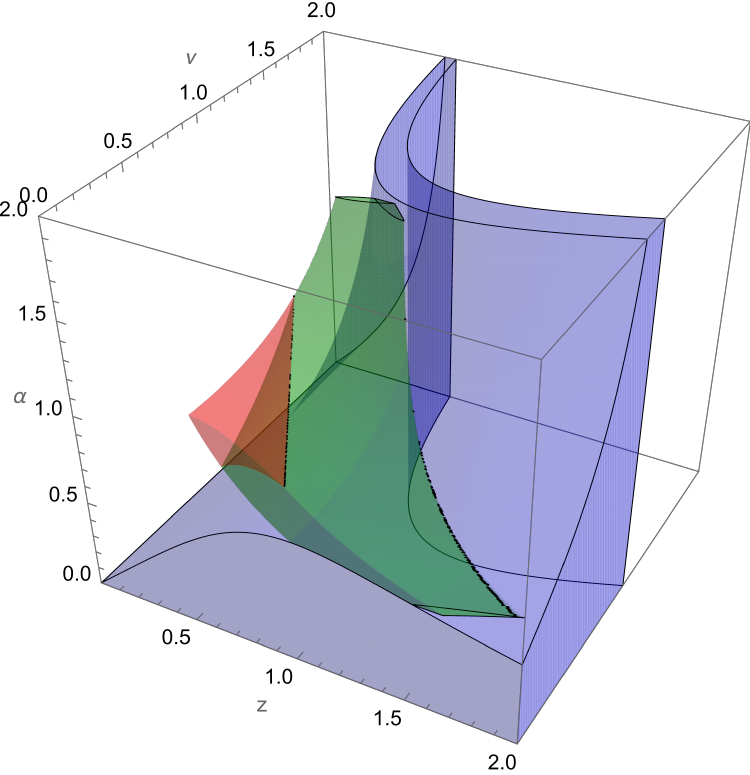}}
			\caption{\label{fig:main}Parameter space for the rotating qBTZ black hole. The blue region corresponds to solutions analogous to classical, rotating BTZ black holes, where $0\leq\alpha\leq\alpha_{\text{ext}}$. The green and red regions correspond to quantum black holes `past extremality', with $\alpha>\alpha_{\text{ext}}$. The latter contains black holes with $\mathcal{R}_{\text{Q}}<1$ but are thermodynamically unstable.}
\end{figure}

For the neutral, static qBTZ, the ratio (\ref{eq:qratio}) was considered in \cite{Frassino:2022zaz} without the subtraction of the Casimir volume. That ratio was found to obey $R_{\mathcal{Q}}>1$ for weak backreaction $\nu\ll1$, but was severely violated for arbitrary $z$ and $\nu$. Further, that ratio is imaginary for dressed conical singularities, as $V_{3}$ is negative. Subtracting the Casimir volume, we find $\mathcal{R}_{\text{Q}}\geq 1$ is provably true for both the neutral and charged qBTZ black holes. For rotating black holes we find potential violations, $\mathcal{R}_{\text{Q}}<1$.  All such violations, however, occur for thermally unstable black holes and are localized to a small region in the $(\alpha, z, \nu)$ parameter space (for rotation parameter $\alpha$) where non-perturbative effects become dominant. See Fig.~\ref{fig:main}.

We find two types of rotating black holes for a given set $(z, \nu)$. The first family of solutions obeys $0\leq\alpha\leq\alpha_{\text{ext}}$, where $\alpha_{\text{ext}}$ is analogous to the classical value for extremality (see supplemental material).
%\be\label{eq:ext} \alpha_{\text{ext}}=\frac{z \sqrt{1+\nu  z}}{\sqrt{1+z^4-2 \nu  z^3}}.\ee
For these black holes, physical quantities such as $T$ or $J$ are monotonic in $\alpha$ as for classical BTZ. A second family of solutions with $\alpha>\alpha_{\text{ext}}$ exists, which is possible because of the combined, non-linear effects of rotation and backreaction. We dub these \emph{nonperturbative rotating black holes}. Among them, it is possible to have $\mathcal{R}_{\text{Q}}<1$, however, all are thermodynamically unstable. Specifically, we find
\be 
C_{V, J, c_3} \equiv T \left(\frac{\partial S_{\text{gen}}}{\partial T} \right)_{V, J, c_3}\!\!<0\,,
\ee
in accord with the conjecture that black holes violating RII (\ref{eq:classRII}) have negative heat capacity at fixed volume \cite{Johnson:2019mdp}.

\noindent \textbf{Discussion.} We proposed semiclassical generalizations of the Penrose and reverse isoperimetric inequalities for asymptotically AdS spacetimes and found they are obeyed for all known quantum AdS$_{3}$ black holes, at any order of backreaction. Our work has many implications and opens other avenues worth exploring.

\noindent\emph{Quantum cosmic censorship.} Historically, the Penrose inequality arose from the search for a counterexample to the weak cosmic censorship conjecture (WCCC) \cite{Penrose:1973um}, i.e.,
%an asymptotically flat (AdS) spacetime with regular initial conditions will be strongly asymptotically predictable. Under cosmic censorship, initial data containing a future trapped surface will lead to a black hole. Penrose further asserts 
singularities lie behind event horizons, out of sight to an observer at future null infinity \cite{Penrose:1968ar,Penrose:1969pc}. As a necessary condition to WCCC, any violation of the Penrose inequality suggests a violation of weak cosmic censorship.
%In the presence semiclassical effects, singularities are still guaranteed to form under gravitational collapse \cite{Wall:2010jtc}.
Black holes that evaporate completely will produce a naked singularity \cite{Kodama:1979vm,Wald:1984rp}. 
%At the endpoint of evaporation, however, the semiclassical approximation breaks down.
This implies the need for a quantum generalization of cosmic censorship. 

In fact, semiclassical gravity predicts quantum cosmic censorship. For example, the classical BTZ geometry with negative mass describes naked conical singularities. Accounting for backreaction of the Casimir stress-tensor, the singularities become shrouded by a horizon \cite{Emparan:2002px,Casals:2016ioo,Casals:2019jfo}. Our work thus represents a consistency between the quantum Penrose inequality and cosmic censorship. As with classical censorship, a reasonable expectation is for the quantum Penrose inequality to be a necessary condition to quantum WCCC. It would be interesting to develop a precise notion of quantum cosmic censorship using the Penrose inequality described here as an input assumption. Further, there is evidence the rotating quantum BTZ black hole obeys strong cosmic censorship \cite{Emparan:2020rnp,Kolanowski:2023hvh}, nor can it be overspun to shed its horizon \cite{Frassino:2024fin}, thus passing a standard test of classical WCCC \cite{Wald:1974hkz,Sorce:2017dst}. It would be worth connecting this gedanken experiment to the quantum Penrose inequality.  
%From the bulk viewpoint this is unsurprising since the four-dimensional black hole spacetime obeys the classical Penrose inequality and WCCC. 

%Semiclassically, the generalized second law of black hole thermodynamics implies a quantum singularity theorem

\noindent \emph{Beyond AdS$_{3}$.} In three-dimensional vacuum general relativity, there are no asymptotically flat or de Sitter black holes. Instead, a point mass in such geometries describes a naked conical singularity \cite{Deser:1983nh,Deser:1983tn}. Black hole horizons induced via backreaction cloak these naked singularities \cite{Souradeep:1992ia,Emparan:2002px,Emparan:2022ijy,Panella:2023lsi}. It is not clear how such censorship is related to a Penrose inequality. Indeed, in the three-dimensional flat context, null shells of dust do not collapse to a black hole, and the classical Penrose inequality is trivial. Further study of three-dimensional asymptotically flat black holes would lend insight into the relation between quantum cosmic censorship and the Penrose inequality.

Descending to one dimension lower, semiclassical backreaction is exactly solvable in dilaton gravity, leading to a host of two-dimensional quantum black holes (cf. \cite{Fabbri:2005mw,Almheiri:2014cka}). Classically, such models arise from the dimensional reduction of, for example, near-extremal black holes, while their semiclassical extension may be realized via braneworld holography \cite{Neuenfeld:2024gta}. In particular, the semiclassical first law (\ref{eq:qbtzfirstlaw}) also holds \cite{Pedraza:2021cvx,Svesko:2022txo}, where the $D$-dimensional black hole area is encoded in the dilaton. For example, for spherical black holes, $A_{\text{BH}}=L^{(D-2)}_{D}\Omega_{D-2}\Phi$, for some relevant $D$-dimensional scale $L_{D}$ and dilaton $\Phi$. Dimensional reduction of the Penrose inequalities (\ref{eq:classAdSPI}) and (\ref{eq:qAdSPI}) thus gives a proposal for their counterparts for two-dimensional dilatonic black holes. It would be worth seeing how these dimensionally reduced inequalities relate to two-dimensional cosmic censorship \cite{Russo:1992yh,Hayward:1992gi,Vaz:1996kh}. 
%\rah{What is meant by the dimensional reduction of the inequality?} \AS{Similar to `dimensional reduction' of entropy, it is plugging in a metric ansatz for dimensional reduction into a higher-dimensional quantity like area.}

It would be most interesting to test our quantum inequalities in higher dimensions, using quantum black holes as a guide. The conjecture of \cite{Emparan:2002px} states braneworld black holes in any dimension map to quantum black holes on the brane. So far, however, finding exact braneworld black holes for $D>4$ has proven difficult. In fact, the few known analytic solutions 
%\cite{} -- found by introducing a pair of branes inside a bulk with a black string -- 
have a vanishing quantum stress-tensor (up to a conformal anomaly), such that the brane geometry appears classical \cite{Fitzpatrick:2006cd,Gregory:2008br}. While this means the classical inequalities for such braneworld black holes will be obeyed, it is unclear if such solutions can be used to test the quantum inequalities.  

\noindent \emph{Beyond Hartle-Hawking states.}  A notable limitation of our tests is that the holographic cutoff conformal field theory is always in the Hartle-Hawking state. Of course, the matter could be in other quantum states, e.g., Boulware or Unruh vacua, and, ideally, the proposed quantum inequalities hold for any Hadamard state. Testing the inequalities for other quantum states beyond perturbative backreaction would thus be of considerable interest. Doing so requires a better understanding of the holographic construction out of equilibrium states, a challenging task (cf. commentary in \cite{Hubeny:2009ru,Marolf:2013ioa}), though progress has been made via numerics \cite{Figueras:2011va} or the large-$D$ approximation \cite{Emparan:2023dxm}.

\noindent \emph{Nature abhors superentropic black holes.} Black holes that violate the reverse isoperimetric inequality are said to be superentropic because their entropy exceeds the amount AdS-Schwarzschild of the same thermodynamic volume would have. All known superentropic black holes have negative heat capacity at constant volume, indicating superentropic black holes are thermodynamically unstable \cite{Johnson:2019mdp} (not all unstable black holes are superentropic, however).  Likewise, we find the only black holes that violate the quantum reverse isoperimetric inequality are thermodynamically unstable.\footnote{Phase transitions and heat capacities of the static qBTZ black hole were examined in \cite{Frassino:2023wpc,Johnson:2023dtf,HosseiniMansoori:2024bfi}. For large backreaction, the qBTZ appeared to be superentropic whilst not necessarily having negative heat capacity. In these articles, however, the Casimir volume was not subtracted.} This tells us that thermodynamic volume is not a mere accident of classical physics; even when accounting for quantum effects the volume remains a diagnostic for thermal stability. Further, discarding the superentropic black holes on physical grounds implies there exists a maximum entropy state for quantum black holes at fixed volume.

%Via AdS$_{3}$/CFT$_{2}$, superentropicity of the charged BTZ black hole can be microscopically understood as the (naive) Cardy formula overcounting the dual CFT$_{2}$ entropy \cite{Johnson:2019wcq}. Since the brane black holes considered here are asymptotically AdS$_{3}$, the induced brane gravity has another dual description in terms of the CFT$_{3}$ coupled to a defect CFT$_{2}$. It would be interesting to see if superentropicity of quantum AdS$_{3}$ black holes has a microscopic herald akin to the classical counterpart. 

%\rah{Maybe need to comment on or at least consider any limitations that may arise due to the particular quantum states defined by holography in this way.}

\noindent \emph{Acknowledgements.}
We are grateful to Roberto Emparan, Eleni Kontou, David Kubiz{\v n}{\' a}k, José Navarro-Salas, Marija Toma\v{s}evi\'{c}, and Jorge Rocha for useful discussions and correspondence.  AMF acknowledges the support of the MICINN grants PID2019-105614GB-C22, AGAUR grant 2017-SGR 754, PID2022-136224NB-C22 funded by MCIN/AEI/ 10.13039/501100011033/FEDER, UE, and
State Research Agency of MICINN through the ‘Unit of Excellence Maria de Maeztu 2020-2023’ award to the Institute of Cosmos Sciences (CEX2019-000918-M).
The work of RAH received the support of a fellowship from ``la Caixa” Foundation (ID 100010434) and from the European Union’s Horizon 2020 research and innovation programme under the Marie Sklodowska-Curie grant agreement No 847648 under fellowship code LCF/BQ/PI21/11830027. JFP is supported by the `Atracci\'on de Talento' program grant 2020-T1/TIC-20495 and by the Spanish Research Agency through the grants CEX2020-001007-S and PID2021-123017NB-I00, funded by MCIN/AEI/10.13039/501100011033 and by ERDF A way of making Europe. AS is supported by STFC grant ST/X000753/1.

\bibliographystyle{apsrev4-2}
\bibliography{qineqrefs}

\newpage

\onecolumngrid

\section{Supplemental material}

\subsection{Bulk and brane set-up}

The starting point of all known exact descriptions of three-dimensional braneworld black holes \cite{Emparan:1999wa,Emparan:1999fd} is the AdS$_{4}$ C-metric. The line element in Boyer-Lindquist-like coordinates (explicitly with a vanishing NUT parameter) is
\begin{equation}
\begin{split}
 ds^2 = \frac{\ell^2}{(\ell + xr)^2} \biggr[ & -\frac{H(r)}{\Sigma(x,r)} \left(dt+ax^2 d\phi \right)^2 + \frac{\Sigma(x,r)}{H(r)} dr^2 + r^2 \left(\frac{\Sigma(x,r)}{G(x)}dx^2 + \frac{G(x)}{\Sigma(x,r)}\left( d\phi- \frac{a}{r^2} dt \right)^2  \right) \biggr] \ ,
 \end{split}
\label{eq:rotatingCmetqbtz}\end{equation}
with
\beq
\begin{split}
    &H(r)= \frac{r^2}{\ell_3^2}+ \kappa -\frac{\mu \ell}{r}+ \frac{a^2}{r^2}+\frac{q^{2}\ell^{2}}{r^{2}}  \ , \quad G(x)= 1-\kappa x^2-\mu x^3+ \frac{a^2}{\ell_3^2}x^4+q^{2}x^{4} \ , \\
   &\Sigma(x,r)= 1 + \frac{a^2 x^2}{r^2} \ . 
\end{split}
\label{eq:metfuncsrotqbtz}\eeq
This geometry may be interpreted as describing a single or pair of uniformly accelerating black holes due to a cosmic string or strut (see, e.g., \cite{Griffiths:2006tk}) with (inverse) acceleration $\ell$. It is a solution to Einstein-Maxwell-AdS$_{4}$ gravity, 
\beq I=\frac{1}{16\pi G_{4}}\int d^{4}x\sqrt{-\hat{g}}\left[\hat{R}+\frac{6}{\ell_{4}^{2}}-\frac{\ell_{\star}^{2}}{4}F^{2}\right]\;, \quad \ell^{2}_{\star}=\frac{16\pi G_{4}}{g_{\star}^{2}}\eeq
where $\ell_{\star}$ is a coupling constant with dimensions of length, $g_{\star}$ is the dimensionless gauge coupling constant, and the AdS$_{4}$ radius $\ell_{4}$ 
\beq \frac{1}{\ell_{4}^{2}}=\frac{1}{\ell_{3}^{2}}+\frac{1}{\ell^{2}}\;.\eeq
Further, while not yet apparent $\kappa=\pm1,0$ corresponds to types of slicings of the boundary ($\kappa=-1$ will ultimately result in BTZ black holes), $\mu$ is a parameter related to the mass of the black hole, non-negative parameter $a$ introduces rotational effects, and $q$ serves as a charge parameter, obeying $q^{2}=e^{2}+g^{2}$ for electric and magnetic charge $e$ and $g$, respectively.

The real zeroes $x_{i}$ of $G(x)$ give rise to conical singularities. These defects result in a cosmic string suspending, say, a single black hole away from the center of AdS resulting in its acceleration. One of these conical singularities can be removed by imposing regularity to ensure smoothness of the geometry along the axis of rotational symmetry, 
\beq \phi\sim \phi+\Delta\phi\;,\quad \Delta\phi=\frac{4\pi}{|G'(x_{i})|}\;.\eeq
In these constructions the smallest positive root, denoted $x=x_{1}$, is chosen, leaving conical singularities at the remaining zeroes $x_{i}\neq x_{1}$. Thus, one restricts themselves to the region $0\leq x\leq x_{1}$, where there are no other conical singularities, and the specific range of $x_{1}$ depends on $\mu,q$ and $a$. Combined with $\kappa$, the $(x_{1},\kappa)$ parameter leads to a family of braneworld black holes \cite{Emparan:2020znc,Climent:2024nuj}.

A notable feature of the C-metric (\ref{eq:rotatingCmetqbtz}) is that the hypersurface $x=0$ is totally umbilic, i.e., the extrinsic curvature $K_{ij}$ is proportional to the induced metric at $x=0$; $K_{ij}=-\ell^{-1}h_{ij}$. Thus, a codimension-1 brane $\mathcal{B}$ placed at $x=0$ is guaranteed to obey the Israel junction conditions. Assuming a purely tensional brane, characterized by the action 
\beq I_{\text{brane}}=-\tau\int_{\mathcal{B}}d^{3}x\sqrt{-h}\;,\eeq
the Israel junction conditions set the tension to be 
\beq \tau=\frac{1}{2\pi G_{4}\ell}\;.\label{eq:branetengen}\eeq
the remaining conical singularities live in the range $x<0$. Further, treating the $x=0$ hypersurface as an end-of-the-world brane, the region $x<0$ (where the remaining conical singularities reside) is cutoff from the rest of bulk AdS$_{4}$. The space can be completed by introducing a second copy of the $0\leq x\leq x_{1}$ region and gluing it along $x=0$, resulting in a $\mathbb{Z}_{2}$-symmetric double-sided braneworld \cite{Randall:1999vf,Karch:2000ct},  without a cosmic string.

Via braneworld holography \cite{deHaro:2000wj}, the induced action on the brane is described by a specific semiclassical higher-derivative theory of gravity coupled to a large-$c$ three-dimensional conformal field theory with an ultraviolet cutoff $\ell$. The theory follows from integrating out the bulk between the AdS boundary and the brane as in holographic regularization \cite{deHaro:2000vlm,Skenderis:2002wp} (see \cite{Emparan:2023dxm} for a pedagogical summary). The precise form of the brane-induced action is not necessary for our purposes but can be found in, e.g., \cite{Emparan:2020znc,Climent:2024nuj}. Relevant for us, however, is that effective couplings on the brane are induced from the higher-dimensional parent theory couplings $\{G_{4},\ell_{4},\ell_{\star},\tau\}$
\bea
G_{3}&=&\frac{1}{2\ell_{4}}G_{4} \,,\label{eq:effGd}\\
\frac{1}{L_{3}^2}&=&\frac{2}{\ell_{4}^2}\left(1-2\pi G_{4}\ell_{4}\tau\right)\,,\\
\tilde{\ell}_{\star}^{2}&=&\frac{5}{4}\ell_{\star}^{2}\;,\quad \tilde{g}_{\star}^{2}=\frac{2}{5}\frac{g_{\star}^{2}}{\ell_{4}}\;.
\label{eq:effLd}
\eea
Here $L_3$ is the effective AdS$_{3}$ radius on the brane and for small backreaction approximately equals the curvature radius $\ell_{3}$. The induced electromagnetic couplings $\tilde{\ell}_{\star}$ and $\tilde{g}_{\star}$ are such that $\tilde{\ell}_{\star}^{2}=\frac{16\pi G_{3}}{\tilde{g}_{\star}^{2}}$. As recognized in \cite{Frassino:2022zaz}, variations solely of the brane tension are reinterpreted as variations of the induced brane cosmological constant, such that mechanical work done by the brane in the bulk perspective yields extended black hole thermodynamics from the brane perspective.

\begin{center}
\textbf{Quantum black holes: geometry and thermodynamics}
\end{center}

We now summarize the geometry and extended thermodynamics of quantum black holes. For convenience, we consider the charged and rotating quantum BTZ black holes separately. 

\vspace{2mm} 

\emph{Charged quantum BTZ}

After imposing bulk regularity, the geometry of the C-metric with an AdS$_{3}$ Karch-Randall brane \cite{Karch:2000ct} at $x=0$  is \cite{Climent:2024nuj,Feng:2024uia} 
\beq
\begin{split}
&ds^2=-f(r)dt^2+\frac{dr^2}{f(r)}+r^2d\phi^2\,,\\
&f(r)=\frac{r^2}{\ell_3^2}-8\mathcal{G}_3M-\frac{\ell F(M,q)}{r}+\frac{\ell^{2}Z(M,q)}{r^{2}}\,,
\end{split}
\label{eq:qBTZ}\eeq
with form functions 
\beq F(M,q)=8\frac{(1-\kappa x_{1}^{2}-q^{2}x_{1}^{4})}{(3-\kappa x_{1}^{2}+q^{2}x_{1}^{4})^{3}}\;,\quad Z(M,q)=\frac{16q^{2}x_{1}^{4}}{(-3+\kappa x_{1}^{2}-q^{2}x_{1}^{4})^{4}}\;.\eeq
The metric is known as the charged quantum BTZ black hole and reduces to the neutral qBTZ \cite{Emparan:2020znc} described in the main text when $q=0$. (See \cite{Emparan:2000fn} for a flat Randall-Sundrum construction \cite{Randall:1999vf}, though the solution was not interpreted as a quantum black hole.)

The (extended) thermodynamic quantities of the quantum black hole are \cite{Climent:2024nuj} (see also \cite{Feng:2024uia}, however, they do not have the same conventions for bulk electromagnetic couplings)
\beq
\begin{split}
&M=\frac{\sqrt{1+\nu^{2}}}{2G_{3}}\frac{z^{2}(1+\nu z)(1+\nu z^{3}(\gamma^{2}-1)+\nu^{2}z^{4}\gamma^{2})}{(1+z^{2}(3+\gamma^{2})+2\nu z^{3}(1+\gamma^{2})+z^{4}\nu^{2}\gamma^{2})^{2}}\;,\\
&T=\frac{1}{2\pi\ell_{3}}\frac{(2+3\nu z+\nu z^{3}(1-\gamma^{2})-2\nu^{2}\gamma^{2}z^{4}-\nu^{3}\gamma^{2}z^{5})}{(1+z^{2}(3+\gamma^{2})+2\nu z^{3}(1+\gamma^{2})+z^{4}\nu^{2}\gamma^{2})}\;,\\
&S_{\text{gen}}=\frac{\pi\ell_{3}\sqrt{1+\nu^{2}}}{G_{3}}\frac{z}{(1+z^{2}(3+\gamma^{2})+2\nu z^{3}(1+\gamma^{2})+z^{4}\nu^{2}\gamma^{2})}\;,\\
&Q_{e}=\sqrt{\frac{16\pi}{5\tilde{g}_{\star}^{2}G_{3}}}\frac{\gamma_{e} z^{2}(1+\nu z)\sqrt{1+\nu^{2}}}{(1+z^{2}(3+\gamma^{2})+2\nu z^{3}(1+\gamma^{2})+z^{4}\nu^{2}\gamma^{2})}\;,\\
&\mu_{e}=\sqrt{\frac{5\tilde{g}_{\star}^{2}}{4\pi G_{3}}}\frac{\nu\gamma_{e}z^{3}(1+\nu z)}{(1+z^{2}(3+\gamma^{2})+2\nu z^{3}(1+\gamma^{2})+z^{4}\nu^{2}\gamma^{2})}\;,\\
&P_{3}=\frac{1}{8\pi L_{3}^{2}G_{3}}=\frac{\sqrt{1+\nu^{2}}}{4\pi \nu^{2}\ell_{3}^{2}G_{3}}(\sqrt{1+\nu^{2}}-1)\;,\\
&V_{3}=-2\pi\ell_{3}^{2}z^{2}(1+\nu z)\frac{(-2-2\nu z+\nu^{2}+2z^{2}\nu^{2}+z^{3}\nu^{3}(1+\gamma^{2})+z^{4}\gamma^{2}\nu^{4})}{(1+z^{2}(3+\gamma^{2})+2\nu z^{3}(1+\gamma^{2})+z^{4}\nu^{2}\gamma^{2})^{2}}\;,\\
&c_{3}=\frac{\ell_{4}^{2}}{G_{4}}=\frac{\ell_{3}\nu}{2G_{3}\sqrt{1+\nu^{2}}}\;,\\
&\mu_{c_{3}}=\frac{z^{2}(1+\nu^{2})\mathcal{N}}{\ell_{3}\nu^{3}(1+z^{2}(3+\gamma^{2})+2\nu z^{3}(1+\gamma^{2})+z^{4}\gamma^{2}\nu^{2})^{2}}\;,
\end{split}
\label{eq:chargeqbtzthermo}\eeq
with $\gamma\equiv qx_{1}^{2}$ and $\gamma_{e}\equiv ex_{1}^{2}$ and
\beq 
\begin{split}
 \mathcal{N}&=4-\nu^{3}z^{3}(5+3\gamma^{2})-2z^{4}\nu^{4}(1+3\gamma^{2})-3\gamma^{2}z^{5}\nu^{5}+\nu z(8+\nu^{2})\\
 &+2(1+\nu z)\sqrt{1+\nu^{2}}(-2-2\nu z+\nu^{2}+2 z^{2}\nu^{2}+\nu^{3}z^{3}(1+\gamma^{2})+z^{4}\gamma^{2}\nu^{4})\;.
 \end{split}
\eeq
The quantities $(M,T,S_{\text{gen}})$ correspond to the mass, temperature, and (classical) entropy of the bulk black hole. Here $Q_{e}$ is the charge associated specific to the electric charge parameter $e$, while $\mu_{e}$ is its associated electric chemical potential; the magnetic charge $Q_{g}$ and chemical potential are given by sending $\gamma_{e}\to\gamma_{g}$. The extended thermodynamic variables include the dynamical pressure $P_{3}=-\Lambda_{3}/8\pi G_{3}$ and the central charge $c_{3}$ together with their conjugate variables, the thermodynamic volume $V_{3}$ and chemical potential $\mu_{c_{3}}$, respectively, and formally defined as
 \beq V_{3}\equiv\left(\frac{\partial M}{\partial P_{3}}\right)_{S_{\text{gen}},c_{3},Q_{e},Q_{g}}\;,\quad \mu_{3}\equiv\left(\frac{\partial M}{\partial c_{3}}\right)_{S_{\text{gen}},c_{3},Q_{e},Q_{g}}\;.\label{eq:formdefV3mu3}\eeq
 In the limit $q\to0$ these variables reduce to those of the static neutral qBTZ black hole \cite{Frassino:2022zaz}, whilst the $(M,T,S)$ are those reported in the main text.

 It is straightforward to verify the thermodynamic quantities (\ref{eq:chargeqbtzthermo}) obey the extended first law
 \beq dM=TdS_{\text{gen}}+\mu_{e}dQ_{e}+\mu_{g}dQ_{g}+V_{3}dP_{3}+\mu_{c_{3}}dc_{3}\;,\eeq
 where variations are computed at fixed values of couplings $G_{3}$ and $g_{3}$. As noted in the main text, pressure variations may be solely induced via variations to the brane tension $\tau$ (\ref{eq:branetengen}). Variations in $c_{3}$, meanwhile, arise from varying either the bulk Newton's constant $G_{4}$ or length scale $\ell_{4}$ (or both).  Further, using a similar scaling argument presented in \cite{Frassino:2022zaz} (as used in Euler's theorem for homogeneous functions), the generalized Smarr relation is 
 \beq 0=TS_{\text{gen}}-2P_{3}V_{3}+\mu_{c_{3}}c_{3}\;.\eeq
 The mass term in the three-dimensional Smarr relation is absent because $G_{3}M$ has a vanishing scaling dimension, as is the case for the classical BTZ black hole \cite{Frassino:2015oca}.

\vspace{2mm}

\emph{Neutral, rotating quantum BTZ}

Upon invoking bulk regularity, the rotating qBTZ black hole at $x=0$ has the line element \cite{Emparan:2020znc} (here we swap notation $r\leftrightarrow\bar{r}$ compared to Eq. (3.15) in \cite{Emparan:2020znc})
\beq
\hspace{-6mm}
\begin{split}
ds^2 = &-\left(\frac{r^2}{\ell_3^2}-8\mathcal{G}_3 M -\frac{\ell \mu \eta^2}{\bar{r}} \right) dt^2 + \left(\frac{r^2}{\ell_3^2}-8\mathcal{G}_3 M + \frac{(4 \mathcal{G}_3 J)^2}{r^2}- \ell \mu (1-\tilde{a}^2)^2 \eta^4 \frac{\bar{r}}{r^2} \right)^{\hspace{-2mm}-1} \hspace{-1mm}dr^2  \\
& + \left(r^2 + \frac{\mu \ell\tilde{a}^2 \ell_3^2 \eta^2}{\bar{r}} \right) d\phi^2 - 8 \mathcal{G}_3 J \left( 1+ \frac{\ell}{x_1 \bar{r}} \right) d\phi d t
\end{split}
\label{eq:qbtzrotatmet}\eeq
where 
\beq \eta=\frac{\Delta\phi}{2\pi}=\frac{2x_{1}}{3-\kappa x_{1}^{2}-\tilde{a}^{2}}\;,\eeq 
with $\tilde{a}\equiv ax_{1}^{2}/\ell_{3}$. Further, 
\beq \bar{r}^{2}\equiv \frac{r^{2}-r_{s}^{2}}{(1-\tilde{a}^{2})\eta^{2}}\;,\qquad r_{s}=2\tilde{a}\ell_{3}\frac{\sqrt{2-\kappa x_{1}^{2}}}{3-\kappa x_{1}^{2}-\tilde{a}^{2}}\;,\eeq
where the $r=r_{s}$ denotes the location of the ring singularity. Here $M$ and $J$ are interpreted as the mass and angular momentum of the braneworld black hole, respectively.

Since it proves important in our analysis of the reverse isoperimetric inequality, let us briefly characterize the family of rotating quantum black hole solutions described in \cite{Emparan:2020znc}. In the neutral, non-rotating case, there are three branches of quantum black holes, branches 1a, 1b, and 2. Branch 1a has $\kappa=+1$ and describes black holes with non-positive mass while branch 1b has $\kappa=-1$ and has non-negative mass black holes, as does branch 2. Branches 1a and 1b smoothly connect to each other (a feature that does not appear for the classical BTZ geometry with positive and negative mass), while branches 1b and 2 meet at an upper bound on the mass. In the case of non-vanishing $J$, there is an analogous set of branches, where, in particular, branches 1b and 2 meet at a maximum value of $M$ for fixed $J$. This occurs when 
\beq x_{1}^{2}+\tilde{a}^{2}=3\;,\quad M=\frac{1}{8\mathcal{G}_{3}}\left(\frac{12}{x_{1}^{4}}-1\right)\;,\quad J=\frac{\ell_{3}}{\mathcal{G}_{3}}\frac{\sqrt{3-x_{1}^{2}}}{x_{1}^{4}}\;.\label{eq:MJextbranch1}\eeq
Notice at $x_{1}=\sqrt{2}$, one attains an extremal bound, where $M=J/\ell_{3}=1/4\mathcal{G}_{3}$. Moreover, among the branch 2 black holes, there is another extremal bound, found by minimizing the mass $M$ for fixed $J$:
\beq \tilde{a}=1\;,\quad M=\frac{J}{\ell_{3}}=\frac{1}{\mathcal{G}_{3}(2+x_{1}^{2})}\;,\label{eq:classextlim}\eeq
which coincides  with the previous extremality bound at $x_{1}=\sqrt{2}$.  Classically, the rotating BTZ black hole obeys the extremality bound $M\geq J/\ell_{3}$. Note, however, for any value of $J$, this classical extremality bound will be violated, $M\leq J/\ell_{3}$, when $-\kappa x_{1}^{2}<2\tilde{a}^{2}$, giving rise to `super-extremal' black holes among the branch 1 solutions.

The standard thermodynamic quantities were reported in \cite{Emparan:1999fd,Emparan:2020znc}. We find the extended thermodynamic quantities to be
\beq 
\begin{split}
&M=\frac{\sqrt{1+\nu^{2}}}{2G_{3}}\frac{(1-\nu z^{3})[z^{2}(1+\nu z)+\alpha^{2}(1+4\nu z^{3}(1+\alpha^{2})-(1+4\alpha^{2})z^{4})]}{[1+3z^{2}+2\nu z^{3}-\alpha^{2}(1+4\nu z^{3}+3z^{4})]^{2}}\;,\\
&T=\frac{1}{2\pi \ell_{3}}\frac{[z^{2}(1+\nu z)-\alpha^{2}(1-2\nu z^{3}+z^{4})][2+3\nu z(1+\alpha^{2})-4\alpha^{2} z^{2}+\nu z^{3}+\alpha^{2}\nu z^{5}]}{z(1+\nu z)[1+\alpha^{2}(1-z^{2})][1+3z^{2}+2\nu z^{3}-\alpha^{2}(1-4\nu z^{3}+3z^{4})]}\;,\\
&S_{\text{gen}}=\frac{\pi\ell_{3}\sqrt{1+\nu^{2}}}{G_{3}}\frac{z(1+\alpha^{2}(1-z^{2}))}{[1+3z^{2}+2\nu z^{3}-\alpha^{2}(1+4\nu z^{3}+3z^{4})]}\;,\\
&J=\frac{\ell_{3}\sqrt{1+\nu^{2}}}{G_{3}}\frac{\alpha z(1+z^{2})[1+\alpha^{2})(1-z^{2})]\sqrt{(1-\nu z^{3})[1+\nu z-\alpha^{2}z(z-\nu)]}}{[1+3z^{2}+2\nu z^{3}-\alpha^{2}(1+4\nu z^{3}+3z^{4})]^{2}}\;,\\
&\Omega=\frac{\alpha(1+z^{2})}{\ell_{3}}\frac{\sqrt{(1-\nu z^{3})[1+\nu z-\alpha^{2}z(z-\nu)]}}{z(1+\nu z)[1+\alpha^{2}(1-z^{2})]}\;,\\
&V_{3}=\frac{2\pi\ell_{3}^{2}\mathcal{N}_{1}}{[1+3z^{2}+2\nu z^{3}-\alpha^{2}(1+4\nu z^{3}+3z^{4})]^{2}}\;,\\
&\mu_{c_{3}}=\frac{(1+\nu^{2})}{\ell_{3}\nu^{3}(1+\nu z)}\frac{\mathcal{N}_{2}}{[1+3z^{2}+2\nu z^{3}-\alpha^{2}(1+4\nu z^{3}+3z^{4})]^{2}}\;,
\end{split}
\eeq
with 
\beq 
\begin{split}
 \mathcal{N}_{1}&=2z^{2}[1+\alpha^{2}(1-z^{2})]^{2}-2\nu z[1-\alpha^{2}(1-z^{2})][\alpha^{2}+2\alpha^{2}z^{4}-z^{2}(2+\alpha^{2})]\\
 &+\nu^{2}\{\alpha^{2}-z^{2}+2\alpha^{2}z^{2}+\alpha^{2}z^{4}[7+2z^{2}+2\alpha^{2}(1-2z^{2}-z^{4})]\}\\
 &-\nu^{3}z^{3}(1+2\alpha^{2})(1+3z^{2}-\alpha^{2}-3\alpha^{2}z^{4})-\nu^{4}z^{6}(1+2\alpha^{2})^{2}\;,
\end{split}
\eeq
and the numerator $\mathcal{N}_{2}$ is cumbersome and not particularly revealing to write down. Here $\alpha\equiv \tilde{a}/\sqrt{-\kappa x_{1}}$. The pressure $P_{3}$ and central charge $c_{3}$ are as given in (\ref{eq:chargeqbtzthermo}), and volume $V_{3}$ and potential $\mu_{c_{3}}$ are defined similarly as in (\ref{eq:formdefV3mu3}), except in terms of fixed angular momentum $J$ instead of fixed charge $Q_{e},Q_{g}$. The extended first law becomes 
\beq dM=TdS_{\text{gen}}+\Omega dJ+V_{3}dP_{3}+\mu_{c_{3}}dc_{3}\;,\eeq
while the Smarr relation is 
\beq 0=TS_{\text{gen}}+\Omega J-2P_{3}V_{3}+\mu_{c_{3}}c_{3}\;.\eeq
When $\alpha=0$ we recover the extended thermodynamics of the neutral, static qBTZ black hole. 

Some restrictions are made when exploring the thermodynamics of the rotating solution \cite{Emparan:2020znc}. In particular, for non-extremal solutions, we take 
\beq 0\leq \alpha^{2}\leq \frac{1+\nu z}{z(z-\nu)}\;.\eeq
The lower bound is chosen such that the background avoids naked closed timelike curves ($\kappa=+1$ solutions have  $\alpha^{2}<0$ and describe negative mass conical defects which have been dressed by a horizon due to backreaction). The upper bound follows from demanding the outer black hole event horizon $r_{+}$ be real and positive.
%, and can be understood as the analog of the classical Kerr bound on how fast a black hole can spin. 
For $\kappa=-1$ and $\nu<z<\nu^{-1/3}$, the upper bound implies $1+\alpha^{2}(1-z^{2})>0$. Even with this bound, however, the temperature $T$ of the black hole can still go negative without further restriction on the range of parameters. Meanwhile, the classical BTZ extremal limit (\ref{eq:classextlim}) occurs when
\beq \alpha^{2}= \alpha^{2}_{\text{ext}}\equiv \frac{z^{2}(1+\nu z)}{1-2\nu z^{3}+z^{4}}\;.\eeq
In this limit the temperature $T$ vanishes.

\begin{figure}[t]
\centerline{\includegraphics[width=0.35\textwidth]{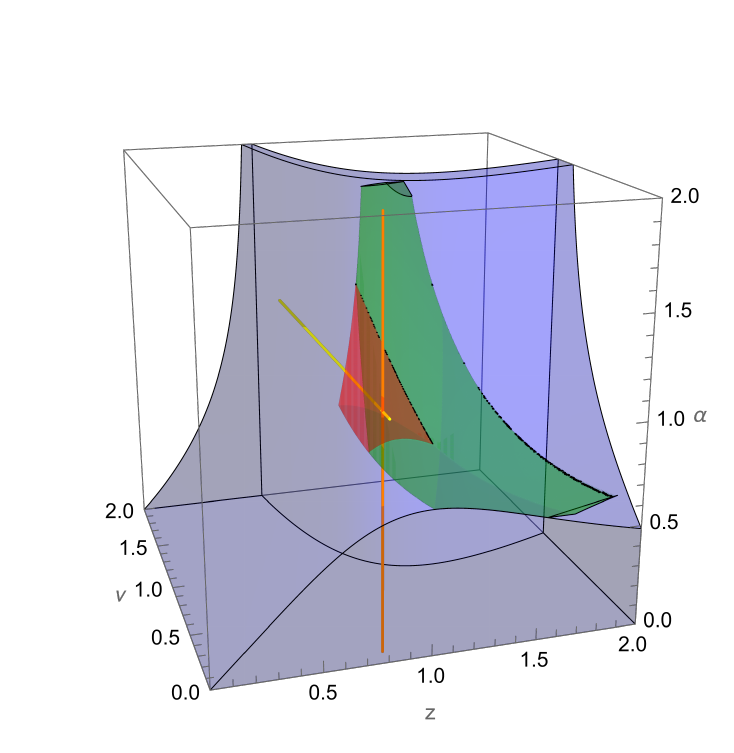}}
			\caption{\label{fig:supp}Parameter space for the rotating qBTZ black hole. The blue region corresponds to solutions analogous to classical, rotating BTZ black holes, where $0\leq\alpha\leq\alpha_{\text{ext}}$, while the green and red regions correspond to quantum black holes `past extremality,' with $\alpha>\alpha_{\text{ext}}$. The latter contains black holes with $\mathcal{R}_{\text{Q}}<1$ but are thermodynamically unstable. Two particular trajectories in the space of parameters are highlighted, the $\alpha$-line and the $\nu$-line, depicted in orange and yellow, respectively.}
\end{figure}

Strikingly, rotating quantum black holes can exist past the classical extremality bound. The reason is that physical quantities such as $T$ or $J$ are, in general, non-monotonic with respect to the rotation parameter $\alpha$. This leads to two distinct types of black holes: i) those respecting $0\leq\alpha\leq \alpha_{\text{ext}}$ and ii) those with $\alpha> \alpha_{\text{ext}}$. In Fig.~\ref{fig:supp} we show a diagram of the allowed range of parameters where we illustrate this point. In this diagram, the blue region corresponds to black holes respecting the bound. Conversely, the green and red regions correspond to `superextremal' solutions. These solutions are only possible due to the combined, non-linear effects of the rotation and quantum backreaction. Hence, they are nonperturbative in nature. 

The red region in Fig.~\ref{fig:supp} contains black holes that violate the RII, namely $\mathcal{R}_{\text{Q}}<1$. However, one can check that these black holes are thermodynamically unstable. To illustrate this, we have depicted two particular trajectories in the space of parameters that go through this region. These are the $\alpha$-line and the $\nu$-line, depicted in orange and yellow, respectively. Some physical data along these two trajectories are shown below in Fig.~\ref{fig:alpha} and Fig.~\ref{fig:nu}. In the first case, we observe a finite gap between `classical' solutions and those that violate the RII. This gap necessarily contains a range of $\alpha$ that is unphysical, i.e., that contains solutions with $T<0$. Likewise, in the second case, we find a range of $\nu$ for which $T<0$. Interestingly, some black hole solutions in the red region seem to have a well-defined $\nu\to0$ limit: classical BTZ black holes with stealth matter. The matter in this case does not backreact on the geometry, however, nevertheless contributes to various thermodynamic quantities. 

\begin{figure}[h]
\centerline{\includegraphics[width=0.33\textwidth]{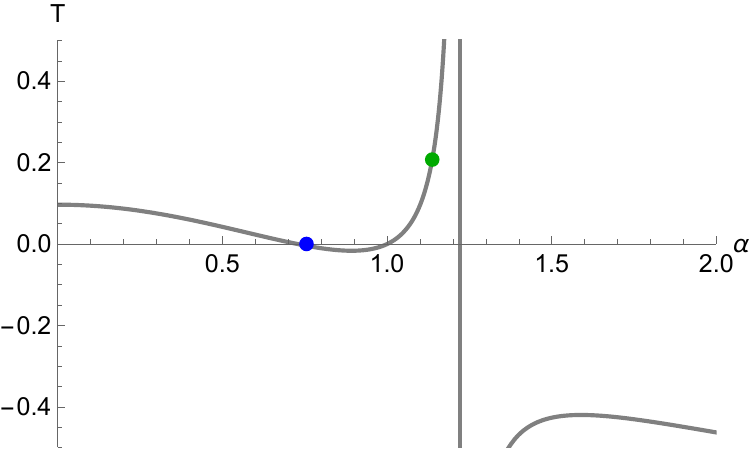}\includegraphics[width=0.33\textwidth]{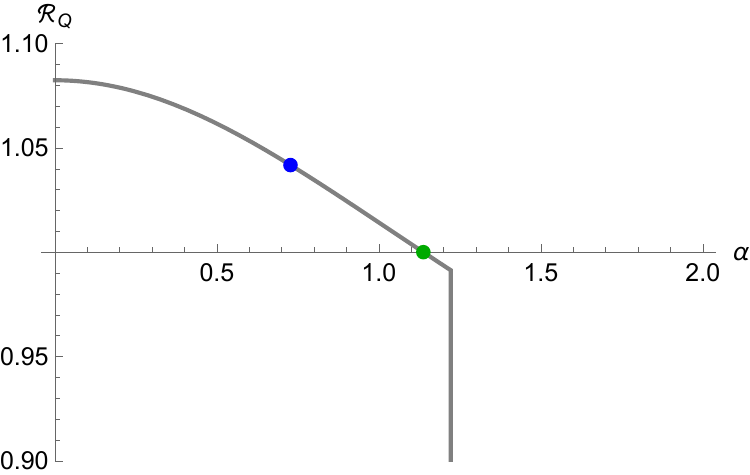}\includegraphics[width=0.33\textwidth]{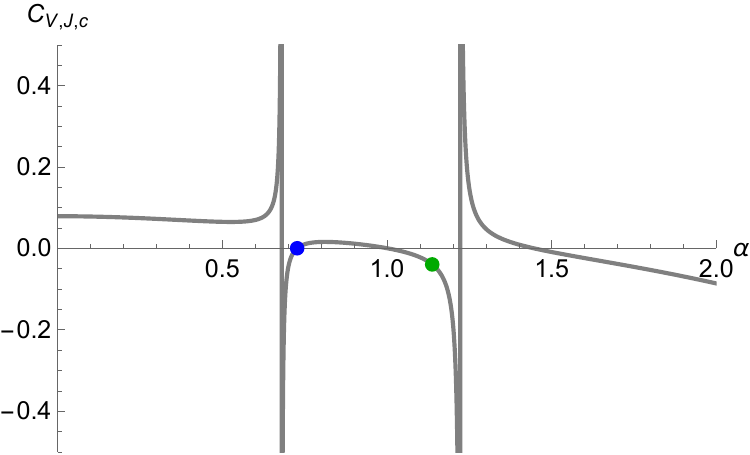}}
			\caption{\label{fig:alpha}Various physical quantities along the $\alpha$-line. For the plots we have set $G_3=1$, $\ell_3=1$, $z=8/10$ and $\nu=1/10$. The blue dot represents the extremal solution with $\alpha=\alpha_{\text{ext}}$ and $T=0$. The green dot is at the interface between the green and red regions, where $\mathcal{R}_{\text{Q}}=1$. It can be checked that $C_{V,J,c}<0$ when $\mathcal{R}_{\text{Q}}<1$ so these solutions are thermodynamically unstable.}
\end{figure}

\begin{figure}[h]
\centerline{\includegraphics[width=0.33\textwidth]{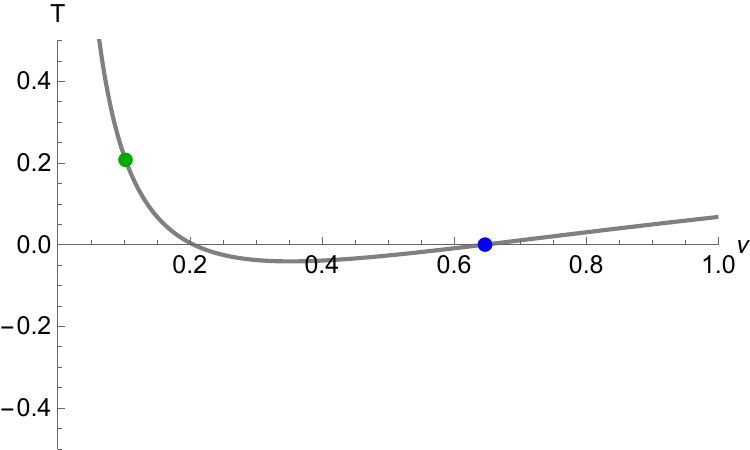}\includegraphics[width=0.33\textwidth]{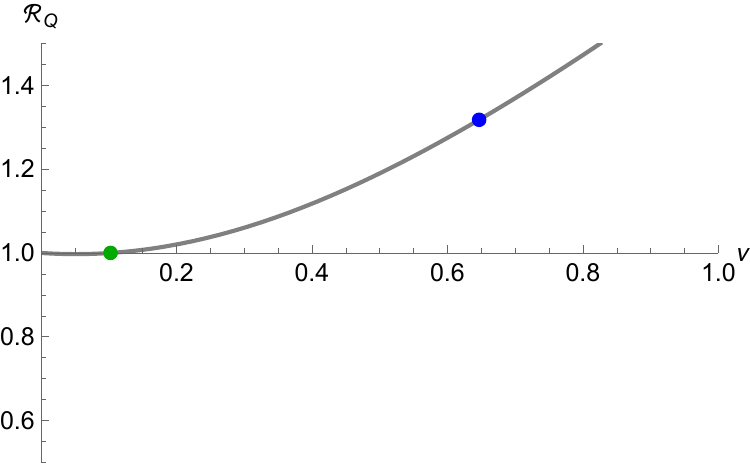}\includegraphics[width=0.33\textwidth]{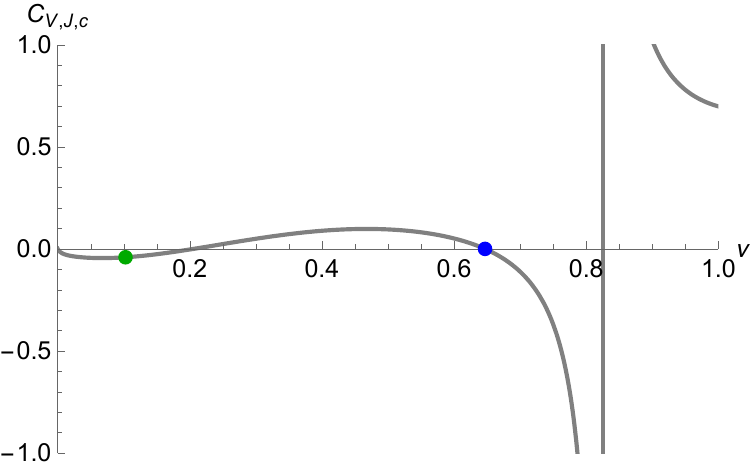}}
			\caption{\label{fig:nu}Some physical quantities along the $\nu$-line. For the plots we have set $G_3=1$, $\ell_3=1$, $z=8/10$ and $\alpha=114/100$. The blue dot represents the extremal solution with $\alpha=\alpha_{\text{ext}}$ and $T=0$. The green dot is at the interface between the green and red regions, where $\mathcal{R}_{\text{Q}}=1$. It can be checked that $C_{V,J,c}<0$ when $\mathcal{R}_{\text{Q}}<1$ so these solutions are thermodynamically unstable.}
\end{figure}

\end{document}